\documentclass[aps,prd,preprint,eqsecnum,nofootinbib,showkeys,%
showpacs,superscriptaddress,a4paper,byrevtex]{revtex4-2}
\usepackage{graphicx}
\usepackage{bm}
\usepackage{amsmath}
\usepackage{amsfonts,amssymb}
\DeclareMathOperator{\R}{Re\,}
\DeclareMathOperator{\I}{Im\,}
\DeclareMathOperator{\Tr}{Tr\,}
\begin{document}
\newcommand{\kb}{k_{\text{B}}}
\newcommand{\mz}{m\hbar\zeta{k_{\text{B}}T}}
\newcommand{\ct}{\coth \left (\frac{\beta \hbar \omega}{2}\right)}
\newcommand{\la}{\langle}
\newcommand{\ra}{\rangle}
\newcommand{\W}{\boldsymbol{\mathfrak{W}}}
\newcommand{\rmi}{\text{i}}
\newcommand{\tm}{T^{\,\text{M}}_{\alpha \beta}}
\newcommand{\ta}{T^{\text{A}}_{\alpha \beta}}
\newcommand{\tsym}{T^{\text{\,sym}}_{\alpha \beta}}
\newcommand{\g}{\ensuremath{\overline\Gamma}}
\newcommand{\wt}{\widetilde{\omega}}
\newcommand{\upl}{e^{-\frac{i}{\hbar}\,Ht}}
\newcommand{\umn}{e^{\frac{i}{\hbar}\,Ht}}
\newcommand{\wrho}{\widetilde{\rho}\,'}
\newcommand{\wA}{\widetilde{A}_j}
\renewcommand{\theenumi}{\roman{enumi}}
\renewcommand{\labelenumi}{\theenumi)}
\def\byrevtex{}
\title{Plasma model and Drude model permittivities \\
in Lifshitz formula}
\author{V.\ V.~Nesterenko}
\email{nestr@theor.jinr.ru}
\affiliation{Bogoliubov Laboratory of Theoretical Physics, 
Joint Institute for Nuclear Research, \\
Dubna 141980, Russia}
\begin{abstract}
At the physical level of rigour  it is shown that there are no 
substantial theoretical arguments in favour of using
either plasma mode permittivity or Drude model permittivity in 
the Lifshitz formula. The decision in this question rests with 
the comparison of theoretical calculations with the experiment. In the 
course of the study the derivation of the fluctuation-dissipation 
theorem is proposed where it is displayed clear at which reasoning 
stage and in what way the dissipation is taken into account. In 
particular it is shown how this theorem works in the case of the 
system with reversible dynamics, that is when dissipation is absent.       
Thereby it is proved that explicit assertion according to which this 
theorem is inapplicable to systems without dissipation is erroneous. 
The research is based on making use of the rigorous formalism of 
equilibrium two-time Green functions in statistical physics at finite 
temperature.

\end{abstract}
\maketitle
\newpage  

\section{Introduction}

The problem of theoretical description of the Casimir forces~\cite{book} 
remains, as before, actual. To a certain extent 
 that is promoted by accumulation of the experimental data in this field.
At~the beginning (Casimir, 1948) these forces were associated with 
the zero point oscillations of electromagnetic field in vacuum~\cite{Milton}. 
Later on (Lifshitz, 1955) a more broad physical base of these forces 
was revealed, namely, its close connection with electromagnetic 
fluctuations in material media was ascertained.  This puts in the 
order of the day correct account of the material characteristics
of the continuous media (first of all, its permittivity and permeability) 
in calculating the Casimir forces and employment here consistent 
mathematical methods (quantum field theory in media, 
fluctuation-dissipation theorem and others). The internal structure of 
the medium proves also to be important~\cite{Fial}.

The substantial difficulties arise in these studies when trying to take 
into account the dissipative properties of the material. A~few years ago
it was found out that the experimental data on the Casimir forces
are described enough good by the Lifshitz formula with the plasma model 
permittivity. An attempt to take into account the dissipation by  
substituting into the Lifshitz formula  the Drude model permittivity 
makes worse  the description of the experimental data 
(see,~e.g.,~\cite{Galina}). These topics were discussed in detail  in many 
publications, we cite here only some of 
them~\cite{Bezerra,Decca,HBAM,Lamb,HB}.  There arose also certain 
difficulties in calculation of the Casimir entropy by making use of  the 
Drude model permittivity \cite{Bezerra,Decca,BorPir,BorEPJ}.

Such a situation turned out to be a surprise  because the Drude model, 
taking account of dissipation, is more realistic in comparison with the 
plasma model which disregards dissipation. This gave  rise to a rather 
long discussion (see papers cited above and references therein).
As  a result it became  clear that it is needed a thorough examination
of the physical and theoretical grounds used in derivation of the Lifshitz 
formula. It is this problem  that is a subject of the present article.

One can distinguish  two basic approaches to obtaining the Lifshitz
formula. First of all  the original approach used by Lifshitz himself 
should be noted~\cite{Lif,LL8-1}. Below we argue that it is based 
substantially on the Callen-Welton fluctuation-dissipation theorem 
or simply FDT \cite{CW,Kubo,Kubo-1,LL5}. Many authors have derived 
the Lifshitz formula by summing up the natural mode contributions to 
vacuum energy of the macroscopic electromagnetic filed in material media 
(the mode-by-mode summation or spectral summation, see, for example, 
\cite{NP} and references therein).  This approach originates in the Casimir 
pioneering paper~\cite{Cas}. Such a summation  wittingly assumes the 
real-valued frequencies. Therefore in this approach the media with 
dissipation cannot be treated because of their complex-valued permittivity, 
for example, the Drude model permittivity. Thus we have to concentrate 
ourselves on analysing the Lifshitz derivation of the formula in question.

Originally \cite{Lif,LL8-1} Lifshitz has obtained his famous formula 
in the framework of the semi-phenomenological fluctuational electrodynamics 
developed by Rytov  \cite{Rytov,RKT}. For the system in equilibrium, 
the Rytov technique amounts to the use of the Callen-Welton 
fluctuation-dissipation theorem \cite{CW,Kubo,Kubo-1,LL5}. 
The proof of this equivalence has been done by Rytov himself  \cite{LR,RKT}  
and by the other authors \cite{Eck,vanHove,Ag}.

In the Lifshitz approach  there is an extremely complicated point, 
namely, the transition to the Matsubara imaginary frequencies $\omega 
\to i \zeta_m, \quad \zeta_m= 2\pi  m({k_{\text{B}}T}/{\hbar} ),\quad 
m=0,\pm 1, \pm 2, \dots $ and $\kb $ is the Boltzmann constant 
(the rotation of the integration path in the complex  $\omega$  
plane). Later on Lifshitz with coauthors \cite{Dz,LL9} repeated 
the derivation of the formula in question by making, from 
the very beginning, use of  the very complicated formalism of quantum 
field theory in the statistical physics~\cite{AGD}.  The central objects 
in this approach  are the Matsubara Green functions  
at finite temperature,  $\mathcal {D}_{ik}(\zeta_n;\mathbf{r_1,r_2})$,  
defined in terms of the retarded Green functions $D^{R}_{ik}(\omega,
 \mathbf{r_1,r_2})$:
\begin{equation}
\label{e0} \mathcal{D}_{ik}(\zeta_n; \mathbf{r_1,r_2})=
D^{R}_{ik}(i\zeta_n,\mathbf{r_1,r_2})\,{.}
\end{equation}
Obviously in this approach the rotation of the integration contour is 
not required, however the equality \eqref{e0} must be proved, in 
other words the transition to the  Matsubara frequencies in the 
retarded Green functions   \eqref{e0} should be substantiated. 
It must be noted that in deriving the Lifshitz formula in Refs.\ 
\cite{Dz,LL9} the FDT is used also.

Thus in different derivations of the Lifshitz formula, which are of interest
for  the aim of the present paper, the employment of the FDT is  a 
substantial step. That is why the main part of the present paper is 
devoted to revealing  the requirements which should be satisfied for 
a correct application of the FDT  when obtaining the Lifshitz formula.
For this purpose, as a preliminary, the FDT is derived in a special way, 
namely, in two stages. At the first stage  the so called  averaged 
anticommutator-commutator relation is obtained  which rigorously 
holds only for the Hamiltonian systems only. 
Then the linear response of the Hamiltonian system  to the external 
action is calculated and the  respective susceptibility is found.
On this basis, as the second stage, the transition to the physical FDT 
is accomplished. Here we clear retrace  in what way the phenomenological 
parameters describing  dissipation are introduced into this theorem. 
Further we  demonstrate conclusively that the requirements  
mentioned above are fulfilled  if the plasma model permittivity 
is used and  show to what extent they are violated when the Drude 
model permittivity is utilized.

The layout of the paper is the following. In Sec.\ II, we derive the 
averaged anticommutator-commutator relation within  the framework 
of the Hamiltonian dynamics. In Sec.\ III, the  linear response theory 
is considered for Hamiltonian systems exerted by external action and 
the respective  susceptibility is obtained. In  Sec.\ IV, the physical 
fluctuation-dissipation theorem is derived and its practical use is 
considered. In Sec.\  V,  we  analyse  the possibility to apply 
the plasma model and Drude model permittivities in the AC-relation 
and in the physical FDT. Preliminary we discuss  the Lifshitz approach 
to description of the electromagnetic field in a material media.  In 
Sec.\ VI, Conclusion, we formulate briefly the main inference  of the 
paper, namely:  there are no theoretical arguments in favour of using 
either plasma model permittivity or the Drude model permittivity in 
the Lifshitz formula.  In Appendix A, we accomplish transition to the 
imaginary frequencies when deriving the Lifshitz formula with the 
plasma model  permittivity at finite temperature.

\section{The averaged anti-commutator-commutator relation for  Hamiltonian systems}
\label{AC-relationL} 

The fluctuation-dissipation theorem~\cite{CW,LL5} connects  the 
quantities characterizing spontaneous equilibrium fluctuations 
(equilibrium correlators) in the system with the generalized 
susceptibility specifying the linear response of the system to 
external action.  As noted in the Introduction we derive  FDT in two 
stages. As the first stage the averaged anticommutator-commutator relation 
(see below)  is proved. We use the formalism of equilibrium two-times 
Green functions  at finite temperature  with real time in statistical 
physics~\cite {Kubo,Kubo-1,ZubUFN-1,Zub-book}.  This formalism 
is based on the Hamiltonian description of the systems under study.

Let us consider a system in thermodynamic equilibrium state at the 
temperature $T$ which is described by a statistical operator, or 
density matrix, $\rho$:
\begin{equation}
\label{e1}
\rho=Q^{-1}\exp(-\beta H), \quad Q=\Tr \exp(-\beta H),
\end{equation}
where $H(q,p)$  is the Hamiltonian of the system and 
$\beta^{-1}=k_{\text{B}} T$. The averaging with respect to the 
equilibrium state will  be denoted by braces
\begin{equation}
\label{e2}
\la \ldots \ra = Q^{-1} \Tr (\rho \ldots )\,{.}
\end{equation}
Let $A_j(q,p)$ is a classical dynamical variable and  $A_j(t)$ is its 
Hermitian operator  in the Heisenberg representation
\begin{equation}
\label{e3}
A_j(t)=e^{\frac{i}{\hbar} H t}A_j(0)\, e^{-\frac{i}{\hbar} H t}{.}
\end{equation}

We remind the derivation  of the so called averaged 
anticommutator-commutator relation (AC-relation from now onwards) 
which expresses the symmetrized correlation function of the operators 
$A_i(t_i),\, A_j(t_j)$
\begin{equation}
\label{e4}
\{
A_i(t_i), A_j(t_j)\}\equiv\frac{1}{2}\,\left  (
\langle A_i(t_i)A_j(t_j)\ra+ \la A_j(t_j)A_i(t_i)\rangle
\right )
\end{equation}
via the averaged values of the retarded commutator of these operators, 
i.e., in terms of the retarded Green function. For the sake of completeness  
we define at once the retarded and advanced Green functions
\begin{gather}
\label{e5} G_{ij}^{\,q} (t)\equiv
\pm\, \frac{1}{i\hbar}\,\theta(\pm \,t)\langle[A_i(t_i),A_j(t_j)]\rangle, \quad
q=r,a, \quad t=t_i-t_j \,{.}
\end{gather}
In this formula   $\theta (t)$ is a step function,
$\theta (t)=1$ for  $t > 0$ and $\theta (t)=0$ at $t < 0$;
the upper sings $(+)$ apply to the retarded Green function, $G^{\,r}(t)$, 
and the lower signs, $(-)$, concern  the advanced Green function, 
$G^{\,a}(t)$.  Below it will be shown that the Green functions 
\eqref{e5}, as well as the pair correlators, depend on the 
deference $t_i -t_j=t$.

Two-time temperature Green's functions \eqref{e5} are a direct 
generalization of the retarded and advanced Green functions introduced  
in quantum field theory (QFT) \cite{BShir}. The distinction is only in 
the averaging method $\langle \ldots \rangle $. In QFT the averaging 
is conducted  with respect to the ground state while  in statistical 
mechanics the Gibbs distribution \eqref{e1}, \eqref{e2} is used 
for this purpose.

From definition  \eqref{e5} we obtain in a straightforward way 
the following relations between the Green functions
\begin{equation}
\label{e5a}
G_{ij}^r(t)=G_{ji}^a(t), \qquad  G_{ij}^a(t)=G_{ji}^r(t)
\end{equation}
and their Fourier transforms
\begin{equation}
\label{e5b}
G_{ij}^r(\omega)=G_{ji}^a(\omega), \qquad  G_{ij}^a(\omega)=G_{ji}^r(\omega)\,{,}
\end{equation}
where
\begin{gather}
G^q_{ij}(t)=\int\limits_{-\infty}^\infty
e^{-i\omega t}G^q_{ij}(\omega)\frac{d\omega}{2\pi}\,{,}\quad  G^q_{ji}(t)=
\int\limits_{-\infty}^\infty
e^{-i\omega t}G^q_{ji}(\omega)\frac{d\omega}{2\pi}\,{;}\nonumber {}\\
G^q_{ij}(\omega)=\int\limits_{-\infty}^\infty
e^{i\omega t}G^q_{ij}(t)\, d t\,{,}\quad  G^q_{ji}(\omega)=
\int\limits_{-\infty}^\infty
e^{i\omega t}G^q_{ji}(t)\,dt\,{,}
\quad q=r,a. \label{e5c}
\end{gather}
In addition it is easy to show that the Green functions \eqref{e5} for the 
Hermitian operators $A=A^\dag $ are real functions of the time~$t$
\begin{equation}
\label{e5d}
\left ( G_{ij}^q(t)\right )^*=G_{ij}^q(t), \quad  q=r,a\,{.}
\end{equation}
From the last two  equations \eqref{e5c} and  definition \eqref{e5} 
it follows, in particular, that the function  $G^r_{ij}(\omega)$
has no singularities in the upper half-plane  of the  complex 
variable $\omega$ and  $G^a_{ji}(\omega)$ in the lower 
half-plane $\omega $.

Further we shall  use the spectral representations of the functions 
under consideration that are closely related with their Fourier 
transforms. Let  $E_\mu$ and $C_\mu$ be the eigenvalues and 
eigenfunctions of the Hamiltonian $H$  in \eqref{e1}:
\begin{equation}
\label{e6}
H C_\mu=E_\mu C_\mu\,{.}
\end{equation}
Obviously,  $C_\mu $ do not depend on $t$. By means of Eqs.\
\eqref{e1}--\eqref{e3} and \eqref{e6} we get the spectral representation 
for the correlator
\begin{equation}
\label{e7} \la A_i(t_i)A_j(t_j)\ra =Q^{-1}\sum_{\nu,
\mu}(C_\nu^*A_i(0)C_\mu)(C_\mu^*A_j(0)C_\nu)\,e^{-\beta
E_\nu}e^{\frac{i}{\hbar}\,(E_\nu-E_\mu)\,t },
\quad t=t_i-t_j\,{.}
\end{equation}
Thus the correlator function $\la A_i(t_i)A_j(t_j)\ra$  and, consequently, 
the Green function \eqref{e5} depend on the deference $t_i-t_j=t$. 
Passing on to the Fourier transform
\begin{equation}
\label{e8} \la A_i(t_i)A_j(t_j) \ra
=\int\limits_{-\infty}^{\infty} e^{-i\omega\, t}
J_{ij}(\omega)\frac{d\omega }{2 \pi}\,{,}
\end{equation}
we deduce from \eqref{e7} and  \eqref{e8}
\begin{gather}
J_{ij}(\omega)=\int\limits_{-\infty}^{\infty} e^{i\omega\,
t}\la A_i(t_i)A_j(t_j) \ra\, dt= \nonumber \\ = 2\pi\, Q^{-1}\sum _{\nu,
\mu}(C_\nu^* A_i(0) C_\mu)(C_\mu^* A_j(0)C_\nu)\, e^{-\beta
E_\nu} \delta \left (\omega +\frac{E_\nu -E_\mu}{\hbar}
\right ){.} \label{e9}
\end{gather}

Doing in the same way  we obtain the spectral representation for the 
correlator with the transposed operators $\langle A_j(t_j)A_i(t_i)\rangle$:
\begin{subequations}
\label{e10} 
\begin{align}
\la A_j(t_j)A_i(t_i)
\ra & =Q^{-1}\sum_{\nu,
\mu}(C_\nu^*A_j(0)C_\mu)(C_\mu^*A_i(0)C_\nu)\,e^{-\beta
E_\nu}e^{\frac{i}{\hbar}\,(E_\mu-E_\nu)\,t }
\label{e10a} \\
& =
Q^{-1}\sum_{\nu,
\mu}(C_\nu^*A_i(0)C_\mu)(C_\mu^*A_j(0)C_\nu)\,e^{-\beta
E_\mu}e^{\frac{i}{\hbar}\,(E_\nu-E_\mu)\,t }, \quad t=t_i-t_j\,{.}
\label{e10b}
\end{align}
\end{subequations}
The last equality  \eqref{e10b} is derived by interchanging the summation 
indexes $\nu \leftrightarrow \mu$ in \eqref{e10a}.

We define the Fourier transform of $\la A_j(t_j)A_i(t_i)\ra$ just as in 
Eq.\  \eqref{e8}
\begin{equation}
\label{e11} \la A_j(t_j)A_i(t_i) \ra = \int
\limits_{-\infty}^{\infty} e^{-i\omega \,t}J_{ji}(\omega)\,
\frac{d\omega}{2 \pi}\,{,} \quad t=t_i-t_j\,{.}
\end{equation}
In view of Eqs.\  \eqref{e11} and  \eqref{e10} we deduce the spectral 
density $J_{ji}(\omega)$
\begin{gather}
J_{ji}(\omega)=\int \limits_{-\infty}^{\infty} e^{i\omega \,
t}\la A_j(t_j)A_i(t_i) \ra \,dt = \nonumber\\ =2\pi Q^{-1}\sum_{\mu,\nu}
(C^*_\mu A_j(0) C_\nu)(C^*_\nu A_i(0) C_\mu)\,e^{-\beta
E_\mu} \delta \left (\omega +\frac{E_\nu -E_\mu}{\hbar}
\right ){.}\label{e12}
\end{gather}
Taking into account  the $\delta$-function in Eq.\  \eqref{e12} we 
can obviously do the substitution
\[
\frac{E_\mu}{\hbar}=\frac{E_\nu}{\hbar}+\omega\,{.}
\]
Comparison of transformed Eq.\  \eqref{e12} with Eq.\  \eqref{e9} 
yields an important relation
\begin{equation}
\label{e13}
J_{ji}(\omega)=e^{-\beta \hbar \omega}J_{ij}(\omega)\,{.}
\end{equation}

This property of the spectral density under consideration is in fact 
a direct consequence  of the trivial equality which is satisfied by 
the density matrix \eqref{e1} and the evolution operator 
$\exp(-\frac{i}{\hbar}\,H\,t)$ in \eqref{e3}  in the case of the 
Hamiltonian systems
\begin{equation}
\label{e14}
\left .\exp\left (
- \,\frac{i}{\hbar}\,H\,t
\right ) \right |_{t=-i\beta \hbar} = Q\,\rho=e^{-\beta H}\,{.}
\end{equation}

With the help of  \eqref{e13}  it is easy to derive the spectral density 
for the symmetrized correlator \eqref{e4}
\begin{equation}
\label{e15}
J_{\{ij
\}}(\omega) =\int \limits_{-\infty}^\infty e^{i\,\omega \,t}
\{
A_i(t_i), A_j(t_j
\}\, dt= \frac{1}{2}\,\left (J_{ij}(\omega)+J_{ji}(\omega)
\right )=\frac{1}{2}\,J_{ij}(\omega)(1+e^{-\beta \hbar \omega})
\end{equation}
and for the commutator $\langle [A_i(t_i), A_j(t_j)] \rangle  $
\begin{equation}
\label{e15a} J_{[ij]}(\omega)= \int
\limits_{-\infty}^\infty e^{i\, \omega \,t} \langle
[A_i(t_i), A_j(t_j)] \rangle \, dt=J_{ij}(\omega
)(1-e^{-\beta \hbar \omega})\,{.}
\end{equation}

The spectral density  $J_{ij}(\omega)$ \eqref{e9} for the Hermitian 
operators $A=A^\dag $ possesses the property
\begin{equation}
\label{e15b}
J_{ij}(\omega)= J_{ji}^*(-\omega)\,{,}
\end{equation}
which can be proved just as the equality \eqref{e13}.
We shall not represent here this trivial proof.

Now we turn to the construction of the spectral density for the Green 
functions \eqref{e5}. For this it will be necessary the integral 
representation of the step function \eqref{e5} (see, for example, 
\cite{ZubUFN-1,Zub-book,BShir}):
\begin{equation}
\label{e16} \theta(t)=\frac{i}{2\pi}\int
\limits_{-\infty}^\infty d\omega
\,\frac{e^{-i\omega\,t}}{\omega+i\varepsilon}\quad  (\varepsilon >0, \;
\varepsilon\to 0)\,{.}
\end{equation}
With the help of Eqs.\ \eqref{e5}, \eqref{e5c}, \eqref{e15a},  and 
\eqref{e16} we get
\begin{equation}
\label{e17} G^q_{ij}(\omega)= \int \limits_{-\infty}^\infty
e^{i\omega\,t} G_{ij}^q (t)\, dt=\frac{1}{2\pi\hbar }\, \int
\limits_{-\infty}^\infty d\omega'
J_{ij}(\omega')\frac{e^{-\beta \hbar
\omega'}-1}{\omega'-\omega \mp i\varepsilon}\,{,} \quad q=r,a\,{.}
\end{equation}
In this formula, as well as in \eqref{e5}, the upper sign $(-)$ 
corresponds to $q=r$ and the lower sign, $(+)$, corresponds to $q=a$.

Previously we have noted that the Green functions for the Hermitian 
operators $A=A^\dag $ are real functions of time (see Eq.\ \eqref{e5d}), 
therefore their Fourier transforms obey the relation
\begin{gather}
\left (G^q_{ij}(\omega)
\right )^*=G^q_{ij}(-\omega), \quad q=r,a {;} \nonumber  \\
{\R} G^q_{ij}(\omega)=
{\R} G^q_{ij}(-\omega) \,{,} \quad {\I} G^q_{ij}(\omega) =
-{\I} G^q_{ij}(-\omega), \quad q=r,a.
\label{e17a}
\end{gather}
This equality can be easily proved by utilizing Eqs.\  \eqref{e17} and 
\eqref{e13}, \eqref{e15b}. Obviously this is a test of consistency of 
the Fourier transform definition  used in Eqs.\ \eqref{e5c}, \eqref{e8}, 
\eqref{e9}, \eqref{e11}, \eqref{e12}, \eqref{e15}, \eqref{e15a}, 
and \eqref{e17}.

For our consideration the case is of a special  interest when
\begin{equation}
\label{e17b}
A_i=A, \quad A_j=A^\dag =A\,{.}
\end{equation}
The point is the AC-relation can be derived in this case easily. 
Indeed, utilizing Eq.\ \eqref{e9} one can show the spectral density 
$J_{AA^\dag}(\omega)$ to be positive
\begin{equation}
\label{e17c}
J_{AA^\dag}(\omega)=2\pi Q^{-1}\sum_{\mu,\nu}\left |(C_\nu^*A(0)C_\mu)
\right |^2 e^{-\beta E_\nu}\delta \left ( \omega +\frac{E_\nu -E_\mu}{\hbar}
\right )>0\,{.}
\end{equation}
By means of the known symbolic equality~\cite{BShir}
\begin{equation}
\label{e17d}
\frac{1}{x\pm i\varepsilon}=  {\cal P}\frac{1}{x} \mp  i \pi
\delta(x), \quad \varepsilon \to + 0
\end{equation}
we deduce from~ \eqref{e17}
\begin{align}
 G^{\, q}_{AA^\dag}(\omega)=&\frac{1}{2\pi \hbar}
\int \limits_{-\infty}^\infty d\omega^\prime
J_{AA^\dag}(\omega^\prime)(e^{-\beta \hbar
\omega'}-1){\cal{P}}\frac{1}{\omega '-\omega } \nonumber \\ &\pm
\frac{i }{2\hbar} J_{AA^\dag }(\omega)(e^{-\beta \hbar \omega}-1),\quad q=r,a.
\label{e17e}
\end{align}
The spectral density $J_{AA^\dag}(\omega)$ is real (see Eq.\ 
\eqref{e17c}), therefore it follows from Eq.\ \eqref{e17e}
\begin{equation}
\label{e17f} \I G^{\,q}_{AA^\dag}(\omega)=\pm \frac{1}{2\hbar}
J_{AA^\dag}(\omega) (e^{-\beta \hbar \omega}-1), \quad q=r,a\,{,}
\end{equation}
By making use of Eqs.\ \eqref{e15}  and \eqref{e17f} we obtain the 
spectral density of the symmetrized correlator $J_{\{AA^\dag\}}(\omega)$:
\begin{equation}
\label{e17g} J_{\{AA^\dag\}}(\omega)=\mp \hbar \, \ct \I
G_{AA^\dag}^{\,q}(\omega),\quad q=r,a\,{.}
\end{equation}
Both signs in the right hand side of   Eq.\ \eqref{e17g} lead to the 
same result because  from  Eq.\ \eqref{e17f} it follows
\[
\I G^{\,r}_{AA^\dag}(\omega)=-\I G^{\,a}_{AA^\dag}(\omega)\,{.}
\]
Therefore we have in the general case
\begin{equation}
\label{e17h}
J_{\{AA^\dag\}}(\omega )= -\hbar \, \ct \I G^{\,q}_{AA^\dag}(\omega){,}\quad q=r,a\, {.}
\end{equation}
Further we shall use the retarded Green function $(q=r)$ to conform 
our consideration to the general principle of causality, namely, 
cause precedes action.

Now we invert definition \eqref{e15}
\begin{equation}
\label{e20} \{ A(t_i), A^\dag (t_j\}=\int
\limits_{-\infty}^\infty e^{-i\,\omega
\,t}J_{\{AA^\dag\}}(\omega)\,\frac{d\omega}{2\pi}\,{,}\quad t=t_i-t_j
\end{equation}
and substitute the spectral density $J_{\{AA^\dag\}}(\omega)$ from 
\eqref{e17h} into \eqref{e20} with $t_i=t_j=\tau$. As a result we get
\begin{gather}
\left .\{
A(t_i), A^\dag (t_j)
\} \right |_{t_i=t_j=\tau}\equiv 
\frac12\,\left (\langle A(\tau)A^\dag(\tau)\rangle +
\langle A^\dag (\tau)A(\tau)\rangle\right )\nonumber \\
=-\frac{\hbar}{2 \pi}\int\limits_{-\infty}^\infty
d \omega \,\ct \I G_{AA^\dag}^{\,r}(\omega)=-\frac{\hbar}{\pi}\int\limits_{0}^\infty
d \omega \,\ct \I G_{AA^\dag}^{\,r}(\omega){.}
\label{e21}
\end{gather}
At the last step in this equation we have taken into account that 
$\I G_{AA^\dag}^{\,r}(\omega)$ is an odd function (see Eq.\ \eqref{e17a}). 
In the first line  of  \eqref{e21} averaging is carried out over the 
system at equilibrium. Hence the result does not depend on time 
$\tau $ and the correlator here can be denoted simply by $\la A^2 \ra$.

Equation \eqref{e21} expresses the symmetrized equilibrium correlator 
of the operators $A(\tau)$ and $A^\dag (\tau)$, Eq.\ \eqref{e4}, 
in terms of the Fourier transform of the retarded Green function, i.e., 
through the retarded commutator of these operators. This formula is 
anticommutator-commutator relation mentioned before Eq.\  \eqref{e4}  
as a preliminary task in deriving FDT. It is important to note that this 
equation is exact and  it is applicable only to the systems that admit 
the Hamiltonian description. In the course of derivation of this 
relation the system has been considered which is at the thermodynamical 
equilibrium.

Theoretically the exact AC-relation \eqref{e21}  may be used in both 
directions, namely, for finding the correlator of equilibrium fluctuations  
through retarded Green function as well as for obtaining this 
function  via the respective correlator. Here it is assumed certainly  
that the Hamiltonian of the system under study is known. However 
the exact calculation of the left hand side and the right hand side 
in AC-relation \eqref{e21} are the tasks of  the same complexity level. 
Therefore the employment of this relation in such calculations does 
not give advantage. These difficulties are  overcome in the physical FDT.

In order to proceed to derivation of the physical FDT  we  need the 
connection of the retarded Green function  with the generalized 
susceptibility. In its turn this requires calculation of the linear 
response of the system under consideration to external action.

\section{The linear response of the Hamiltonian system \\ to external action}
\label{response}
In the preceding Section  we have derived the AC-relation Eq.\  \eqref{e21} 
considering equilibrium  system at the temperature $T$ which is described 
by the Hamiltonian $H$ and the statistical operator $\rho$ (see Eqs.\ 
\eqref{e1} and \eqref{e2}). Now we find the linear response of this 
system to the external action \cite{Kubo,Kubo-1,ZubUFN-1,Zub-book}. 
To this end we add to the Hamiltonian $H$  the following term
\begin{equation}
\label{e25}
-\sum_j A_j F_j(t)\,{,}
\end{equation}
where  $A_j$ are, as before,  the dynamical variables (operators) 
pertaining to the initial system and $F_j (t)$ are non-operator real 
functions describing external forces or fields. It is assumed that 
the fields $F_j(t)$ are adiabatically turned on in the remote past
\begin{equation}
\label{e26}
F_j(t) \sim e^{\varepsilon t}, \quad t\to -\infty ,\quad \varepsilon>0\,{.}
\end{equation}
Thus we are considering the system with a complete Hamiltonian
\begin{equation}
\label{e26a}
H'_t=H -\sum_j A_j F_j(t)\,{.}
\end{equation}
In this equation the subscript $t$ by the operator $H'_t$ points to 
the explicit dependence  of this Schr\"odinger operator on time due 
to  the external fields. The corresponding statistical operator 
$\rho '(t)$ is determined by the equation
\begin{equation}
\label{e27}
i \hbar \,\frac{\partial \rho'(t)}{\partial t}=[H'_t,\rho'(t)]
\end{equation}
with the initial condition
\begin{equation}
\label{e28}
\left .\rho'(t)\right |_{t\to-\infty}=\rho\,{,}
\end{equation}
where  $\rho$ is the equilibrium statistical operator \eqref{e1}.
All the operators, $H, A_j,  H'_t, \rho $, and $\rho'(t)$,  are in 
the Schr\"odinger representation. In the general case  the statistical 
operator $\rho '(t)$ in the Schr\"odinger representation depends 
on time due to construction (see, for example, Ref.\ \cite{blum}).

Now we fulfil the unitary transformation defined by the formulae
\begin{align}
\rho'(t) & =\upl \widetilde{\rho}\,'(t)\, \umn\,{,} \nonumber \\
A_j & = \upl \widetilde{A}_j(t)\, \umn \,{.}
\label{e29}
\end{align}
The parameter $t$ in transformation  \eqref{e29}
is chosen to be equal to the time $t$ in the functional dependence  
$F_j(t)$ in \eqref{e25}.

Strictly speaking the transformation \eqref{e29} implements 
transition to the interaction representation for the system described 
by the Hamiltonian \eqref{e26}. However in the preceding Section 
we have defined the Heisenberg operators $A_j(t)$ by the same 
formulae (see  Eq.\  \eqref{e3}). In Eq.\ \eqref{e29} the use of the 
tilde is required, in fact, only for the statistical operator $\wrho$ 
because this operator depends on $t$ in the Schr\"odinger 
representation too. For brevity and when this does not lead to 
misunderstanding we shall call all the operators, depending on $t$, 
the Heisenberg operators and use for them the notations 
$A_j(t)\equiv\widetilde{A}_j(t)$  on the same footing.

The  equation for $\wrho(t)$ follows from Eq.\  \eqref{e27} with 
allowance for \eqref{e29}
\begin{equation}
\label{e30}
i \hbar \,\frac{\partial \wrho (t)}{\partial
t} = \sum_j [\wrho (t), \widetilde{A}_j(t)] F_j(t)\,{.}
\end{equation}
Here we have taken into account the sign minus in Eq.\ \eqref{e25}.
The initial condition  for $\wrho (t)$ is obtained from Eq.\ \eqref{e28}
\begin{equation}
\label{e31}
\left .\wrho(t) \right |_{t\to-\infty}= \widetilde{\rho} =\rho\,{.}
\end{equation}
Equation \eqref{e30} and condition \eqref{e31} can be united in a single 
integral equation
\begin{equation}
\label{e32} \wrho (t)= \rho +\frac{1}{i\hbar}\int\limits
_{-\infty}^t \sum_j\,[\wrho(\tau),\wA(\tau)]\,F_j(\tau)\, d\tau\,{.}
\end{equation}
For the statistical operator  in the Schr\"odinger representation, 
$\rho'(t)$, we obtain from \eqref{e32}
\begin{equation}
\label{e33}
\rho '(t) = \rho +\frac{1}{i\hbar} \int
\limits_{-\infty}^t \sum _j \,[\wrho(\tau -t),
\wA(\tau -t) ] F_j(\tau)\,d\tau\,{.}
\end{equation}
Till  now our consideration was exact. Further we suppose that the 
external action \eqref{e25} is sufficiently weak, so that the method 
of successive  approximation is applicable and we can confine ourselves 
to the first iteration, i.e.\ substitute $\rho$ for $\wrho(\tau -t)$ in the 
right hand side of \eqref{e33}
\begin{equation}
\label{e34} \rho '(t) \simeq \rho +\frac{1}{i\hbar} \int
\limits_{-\infty}^t \sum _j \,[\rho, \wA(\tau -t) ]
F_j(\tau)\,d\tau\,{.}
\end{equation}

We define the response of the system to external action in the usual 
way \cite{Kubo,Kubo-1,ZubUFN-1,Zub-book}
\begin{gather}
\Delta A_i(t) \equiv \Tr (\rho' (t)A_i)-\la A_i \ra \nonumber \\
= \frac{1}{i\hbar } \int\limits_{-\infty}^t \sum_j\Tr
([\rho,\widetilde{A}_j(\tau-t)]A_i)F_j(\tau) \, d\tau =
\frac{1}{i\hbar } \int\limits_{-\infty}^t \sum_j
\la [\widetilde{A}_j(\tau-t),\widetilde{A}_i(0)] \ra \, F_j(\tau)\, d\tau  \,{.}\label{e35}
\end{gather}
Here we have taken advantage of the equality $A_i=\widetilde{A}_i(0)$ 
that follows from \eqref{e29}. It is worthy to remind that the  braces  
$\la \ldots \ra$ mean, as before, the averaging with respect to 
equilibrium state of an isolated system with statistical operator 
$\rho$ (see Eq.\ \eqref{e2}). With the help of the  
$\theta$-function we firstly extend  the integration 
in \eqref{e35} to $+\infty$ and  then use the  definition of the 
retarded Green function~\eqref{e5}
\begin{align}
\label{e36}
\Delta A_i(t)&=-
\frac{1}{i\hbar } \int\limits_{-\infty}^\infty  \theta(t-\tau)
\sum_j\la [\widetilde{A}_i(0),\widetilde{A}_j(\tau-t)] \ra \, F_j(\tau)\,d\tau  \nonumber \\
&=- \sum_j \int\limits_{-\infty}^\infty G^{\,r}_{ij}
(t-\tau)\,F_j(\tau)\,d\tau = - \sum_j \int\limits_{0}^\infty G^{\,r}_{ij}
(t)\,F_j(t-\tau)\,d\tau\, {.}
\end{align}
Thus the retarded Green function $G^{\,r}_{ij}(t)$ completely defines 
the response of the system with the Hamiltonian $H$ to the external 
influence \eqref{e25} which is taken into account in the  linear 
approximation. 

Now we compare  Eq.\  \eqref{e36} with the definition of the generalized 
susceptibility. For simplicity we beforehand rewrite  Eq.\  \eqref{e36} 
for the case  when sum in Eq.\  \eqref{e25} contains only one 
term and use the   substitution \eqref{e17b}
\begin{equation}
\label{e37}
\Delta A(t)=-\int \limits_{0}^{\infty} G_{AA^\dag}^{\,r}(\tau ) F(t-\tau) d\tau
\,{.}
\end{equation}
In this case the generalized susceptibility $\alpha (\tau)$ is defined 
in the following way (see, for instance, Ref.\ \cite[\S 123, 
Eq.\ (123,2)]{LL5}:
\begin{equation}
\label{e38}
\Delta A(t)=\int \limits_{0}^{\infty} \alpha (\tau ) F(t-\tau) d\tau
{.}
\end{equation}
It is obvious  that in our consideration $\bar x(t)=\Delta A(t)$. 
Juxtaposing equalities \eqref{e37} and \eqref{e38} we infer that the 
retarded Green function taken with the sign minus, 
$-G_{AA^\dag}^{\,r}(\tau ) $, is the generalized susceptibility for 
the systems in question, i.e.\ for the Hamiltonian systems. In 
other words this statement holds only  for systems with reversible 
dynamics. 

\section{Fluctuation-dissipation theorem and its practical use}
\label{FDT}
In  previous  Section it was ascertained the relationship of the 
retarded Green function entering the  AC-relation \eqref{e21}  
with the generalized susceptibility $\alpha (\tau )$ defined in Eq.\ 
\eqref{e38}. Proceeding from this we can modify  the AC-relation in 
such a way that its field of applicability extends beyond the  
Hamiltonian systems including also  slightly irreversible systems. 
For that we simply replace the retarded Green function 
$-G^{\,r}_{AA^\dag}(\omega ) $ in the AC-relation \eqref{e21} by 
the generalized susceptibility $\alpha (\omega)$
 \begin{equation}
\label{e39}
\alpha (\omega)=\int \limits_{0}^{\infty} \alpha (\tau ) e^{i\omega \tau}  d\tau
\end{equation}
(see, for example, Ref.\ \cite[\S 123, Eq.\ (123,4)]{LL5}). As a result we get
 the  physical formulation of the FDT
 \begin{equation}
\label{e39a}
\la A^2
\ra =\frac{\hbar}{2}\int\limits_{0}^\infty
\I \alpha (\omega) \ct d\omega
\end{equation}
(see Ref.\ \cite[\S 124, Eq.\ (124,10)]{LL5}). From the physical point of 
view it is clear that such a substitution is permissible when the  
generalized susceptibility $\alpha (\omega )$ is calculated
for the system with a reversible dynamics (i.e.,  for the Hamiltonian  
system or for the system with slightly irreversible dynamics 
that is for slightly dissipative system. Herewith
in the latter case it is possible  to anticipate that the physical FDT 
Eq.\ \eqref{e39a} will hold with the accuracy proportional
to the extent  of irreversibility or dissipation. In the next 
Section~V it will be demonstrated by making use of the Drude 
model permittivity.

Calculation of $\alpha (\omega)$  for slightly dissipative system 
and its  substitution  into  \eqref{e21} for $G_{AA^\dag}^{\,r}(\omega)$ 
results in appearance in the FDT  \eqref{e39a} of the phenomenological 
parameters responsible for dissipation. This is to be noted in view of 
the following reason.

Practically at once after appearance of the Callen-Welton paper the attempts  
have begun with the aim to answer the question: whether the FDT takes 
into account the dissipation and if yes then in what way. Here are a 
few  typical papers on this subject \cite{Ginz}.  As far as we know a 
complete clearness in this problem is absent till now.

All this is in conformity with the results  of a consistent application of the 
exact AC-relation \eqref{e21} when the correlator  or the Green 
function are obtained by making use of the solutions to the pertinent 
equations generated by the initial Hamiltonian 
$H$~\cite{ZubUFN-1,Zub-book}.  These equations are infinite sequences  
of coupled equations containing the correlators and the Green functions 
of arbitrary high order. The lack of regular procedure for decoupling 
theses equations or cut-off them does not, in particular, allow one 
to evaluate  the accuracy of the obtained results \cite{ZubUFN-2}. 
In this approach nobody has  succeeded in convincing  demonstration 
of a consistent description of dissipation effects starting from the 
Hermitian Hamiltonians
(see the Table at the end of Ref.\ \cite{ZubUFN-1}). Of course, in 
this approach it is impossible to calculate the parameters describing 
dissipation.

\section{Use of plasma model and Drude model permittivities \\
 in AC-relation and in FDT}

At first it is worthy to recall  briefly  in what way the electromagnetic 
field  in a material medium is described in the Lifshitz 
approach~\cite{LL9}.  Here the central point is the use of the  
classical Green function for the macroscopic Maxwell equations as  
the quantum two-time Green function  of the photon  in a medium at 
finite temperature.  The physical argument for this is the fact that 
in this case the linear response of electromagnetic field in medium 
to external action will certainly satisfy the classical macroscopic 
Maxwell equations.

This assertion becomes evident if we remind the fact that the retarded 
Green function is, up to sign, the generalized susceptibility, which 
describes the linear response  of the system to external action (see 
Sec.\ III).  A notable  advantage of the approach in question is that 
it provides an opportunity to avoid the explicit quantization of 
electromagnetic field in a medium, i.e.,  there is no need to settle 
the Hamiltonian, to postulate the canonical commutation relations and 
so on~\cite{Ag,Eck}.

The weak features of this approach are to be noted too. It is obvious 
that one can judge its applicability to a given problem only after 
comparison of the calculations with the experiment. Further in the 
Lifshitz approach the photon Green function in medium at finite 
temperature turns out to be independent of the temperature.\footnote{As 
far as we know there is no admissible explanations of this fact.} In 
this approach the temperature dependence  emerges only due to 
the fluctuation-dissipation theorem.

Now we ascertain whether it is possible to use the plasma model and 
Drude model permittivities in the exact AC-relation \eqref{e21} 
and in the physical FDT \eqref{e39a}. These permittivities appear in the 
stated equations via the classical Green functions of the macroscopic 
Maxwell equations. As it was shown in the Introduction the inferences 
obtained here will be true for  application of the considered  
permittivities in the Lifshitz formula also.

In the Lifshitz treatment of the electromagnetic field in a medium the 
FDT  is used in the field version (see Eq.~(76.6) in Ref.~\cite{LL9}). For the 
sake of simplicity and for a direct connection with our calculations in 
preceding Sections we consider, in place of quantum-field retarded Green 
function
\begin{equation}
\label{e40} D^{\,\mathrm{R}}_{ij}(\omega,
\mathbf{k})=\frac{4\pi \hbar}{\omega^2
\varepsilon(\omega)/c^2-k^2}\left(\delta_{ik}- \frac{c^2
k_ik_j}{\omega^2\varepsilon(\omega)} \right ){,}
\end{equation}
the quantum-mechanical analog of \eqref{e40} omitting the dependence 
on~$\mathbf{k}$:
\begin{equation}
\label{e41} 
G^{\,r}(\omega) = \frac{g}{\omega^2 \varepsilon(\omega)}
\end{equation} 
(further not to be confused the dielectric permittivity 
$\varepsilon(\omega)$ with the infinitesimal positive 
quantity $\varepsilon$.  In Eq.\  \eqref{e41}  $g$ is a positive 
constant the explicit form of which is irrelevant  in what 
follows.\footnote{In the field version of FDT \cite[Ch.~VIII]{LL9}
the variables $\mathbf{r}_i, \mathbf{r}_j$ and, consequently, 
$\mathbf{k}_i, \mathbf{k}_j$ are the continuous counterparts 
of the indices $i,j$ [see Eq.~\eqref{e21}]; hence 
$\mathbf{k}=\mathbf{k}_i- \mathbf{k}_j=0$.} Evidently such a 
simplification of the Green function preserves the time dependence
in the problem at hand in a correct way. In preceding Sections it was 
shown that it is the analysis of this dependence that enables one to 
get the AC-relation in the form~\eqref{e21}.

The substitution of the dielectric permittivity calculated in the plasma model
\begin{equation}
\label{e42} 
\varepsilon_{\text{pl}}(\omega)=1-\frac{\omega^2_{\text{pl}}}{\omega^2}
\end{equation}
for $\varepsilon(\omega)$ in Eq.\ \eqref{e41} gives the respective Green function
\begin{gather}
 G_{\text {pl}}^{\,r}(\omega) =\frac{g}{\omega^2
\varepsilon_{\mathrm{pl}}(\omega)} = \frac{g}{\omega^2
-\omega^2_{\mathrm{pl}}} \nonumber \\
{} = \frac{g}{2\omega_\text{pl}}\left (
\frac{1}{\omega -\omega_{\text{pl}}+i\varepsilon}-\frac{1}{\omega
+\omega_{\text{pl}}+i\varepsilon} \right ) {.} \label{e43}
\end{gather}
In order that to get in Eq.~\eqref{e43} the Fourier transform of 
the retarded Green function we explicitly indicated, at the final step, 
the bypass rules for the poles substituting, as usual, 
$\omega +i \varepsilon,\; \varepsilon>0$ for $\omega$  
(see Eqs.~\eqref{e16}, \eqref{e17}, and Fig.~1, left panel). 
By making use of the symbolic equality~\eqref{e17d} we get the 
imaginary part of the Green function~$G_{\mathrm{pl}}^{\,r}(\omega)$:
\begin{equation}
\label{e44} 
	\I G^{\,r}_{\text{pl}}(\omega) = \frac{\pi g }{2\omega _{\text{pl}}}
[\delta (\omega+\omega_{\text{pl}})-\delta (\omega-\omega_{\text{pl}})]\,{.}
\end{equation}
It makes sense to remind  that equality~\eqref{e17d} can be used 
only for real variable~$x$. Evidently, Eq.~\eqref{e43} meets this 
requirement.

Having obtained $\I G^{\,r}_{\mathrm{pl}}(\omega)$ we can get, utilizing 
Eq.~\eqref{e17f} with $q=r$, the spectral density of the 
correlator $J_{AA^{\dag}}(\omega)$:
\begin{align}
J_{AA^{\dag}}(\omega)&=\dfrac{2\hbar}{e^{-\beta \hbar \omega}-1} 
\I G^{\,r}_{\mathrm{pl}}(\omega)
 \nonumber \\&=
\dfrac{\pi g\hbar}{\omega_{\text{pl}} (e^{-\beta \hbar \omega}-1)}\left [
\delta(\omega+\omega_{\text{pl}})-\delta(\omega-\omega_{\text{pl}})
\right] \nonumber \\
& =\frac{\pi g \hbar}{\omega_{\text{pl}}}\left [
\dfrac{\delta(\omega+\omega_{\text{pl}})}{e^{\beta \hbar \omega_{\text{pl}}}-1} 
- \frac{\delta(\omega-\omega_{\text{pl}})}{e^{-\beta \hbar
\omega_{\text{pl}}}-1} \right ]  >0\,{.}
\label{e44a}
\end{align}
Here we have taken into account the evident inequalities 
$e^x-1>0$ and $e^{-x}-1<0$ which hold for 
$x=\beta \hbar \omega_{\text{pl}} >0$.

Thus the Green function \eqref{e43} with the plasma model permittivity 
complies with requirements which emerge in obtaining such a Green 
function in the Hamiltonian approach, namely, 
$\I G^{\, r}_{\mathrm{pl}}(\omega)$ is a real function and 
$J_{AA^{\dag}}(\omega)$ is a positive function of frequency~$\omega$.

Hence  in the case of the plasma model permittivity the use of the 
AC-relation \eqref{e21} and consequently the use of FDT  \eqref{e39a} 
are completely justified. As shown in the Introduction it is implied 
that the use of this permittivity in the Lifshitz formula is justified too. 

Thus we have here an obvious example of a correct application of FDT 
in the case when the dissipation is absent ($\varepsilon_{\text{pl}}(\omega)$ 
in~\eqref{e42} is a real function).  So the name itself of 
equality~\eqref{e39a}, the fluctuation-dissipation theorem,  is rather 
conditional. Thereby it is proved also that explicit assertion according 
to which   this theorem is inapplicable to systems without
dissipation is erroneous~\cite{Shapiro}. 

Exactly to elucidate this point we have divided the derivation of FDT 
in two steps: at first the AC-relation \eqref{e21} was obtained  and 
after  that the FDT \eqref{e39a}.

Here the following note should be taken also.
In our paper \cite{NP} we have rigorously derived the Lifshitz 
formula at zero temperature by making use of the plasma model 
permittivity. In the present article, in the Appendix~A, we do the 
same at finite  temperature by utilizing  the  integration contours 
such as in \cite{NP} without using FDT. By the way,  as far as 
we know Lifshitz himself, when calculating the Casimir force, 
has employed only real dielectric permittivity~$\varepsilon(\omega)$.

Now we turn to the Drude model permittivity
\begin{equation}
\label{e45}
\varepsilon_{\mathrm{D}}(\omega)=1-\frac{\omega^2_{\mathrm{pl}}
}{\omega
(\omega+2i\gamma)}\,{.}
\end{equation}
For the sake of  formula simplification  in what follows we have 
denoted the relaxation parameter in \eqref{e45} by $2 \gamma$.

The substitution  of \eqref{e45} in  \eqref{e41} gives the retarded 
Green function with the Drude model permittivity
\begin{equation}
\label{e46}
G_{\mathrm{D}}^{\,r}(\omega)=G_1(\omega)+G_2(\omega)\,{,}
\end{equation}
where
\begin{gather}
\label{e47}
G_1(\omega)=\frac{g}{2\wt_{\text{pl}}} \left (
\frac{1}{\omega -\wt_{\text{pl}}+i\gamma} - \frac{1}{\omega +\wt_{\text{pl}}+i\gamma}
\right ){,} \\
G_2(\omega)=-\frac{2i\gamma g}{\omega^2_{\text{pl}}}\cdot 
\frac{1}{\omega+i\varepsilon}+
\frac{i\gamma g}{\wt_{\text{pl}}(\wt_{\mathrm{pl}}+
i\gamma)}\cdot \frac{1}{\omega
+\wt_{\mathrm{pl}}+i\gamma}+
\frac{i\gamma g}{\wt(\wt_{\mathrm{pl}}
-i\gamma)}\cdot \frac{1}{\omega -\wt_{\mathrm{pl}}+i\gamma}\,,
\label{e48} \\
\wt^2_{\mathrm{pl}}=\omega_{\text{pl}}^2-\gamma^2{.}
\label{e49}
\end{gather}
Decomposition  \eqref{e46}  is made in such a way that when 
$\gamma \to 0$ the first term, $G_1(\omega)$, turns into 
$G_{\mathrm{pl}}^{\,r}(\omega)$ and the second term vanishes 
$G_2(\omega)\to 0$.

\begin{figure}[t] 
\label{fig1}
\includegraphics[width=0.47\textwidth,clip]{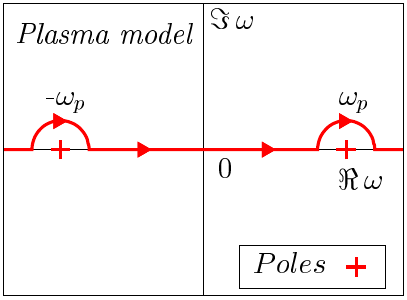} 
\hfill 
\includegraphics[width=0.47\textwidth,clip]{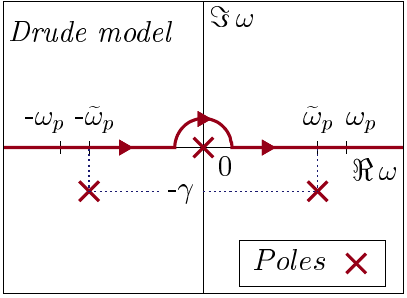} 
\caption{\label{Fig1}(Color online)
The poles of the quantum-mechanical Green functions
$G^{\,r}_{p}(\omega)$ (Eq.\ \eqref{e43}; left panel) and 
$G^{\,r}_{\text{D}}(\omega)$  (Eqs.\ \eqref{e46}--\eqref{e49}; 
right panel). The parameters
$\omega_{\text{pl}},\; \widetilde \omega_{\text{pl}}$, 
and $\gamma$ are specified in Eqs.\ \eqref{e42}, \eqref{e49}, and 
\eqref{e45}, respectively. The thick lines with arrows show the  
paths for integration over $\omega$ in Eqs.\ \eqref{e5c}.}
\end{figure}

The analytical properties  of the Green functions  
$G_{\mathrm{D}}^{\,r}(\omega)$ and  
$G_{\mathrm{pl}}^{\,r}(\omega)$
prove to be substantially different. According to \eqref{e46}--\eqref{e49} 
the Green function $G_{\mathrm{D}}^{\,r}(\omega)$ has three poles 
(Fig.~\ref{Fig1}, right panel). Two of them are located  in the lower half-plane 
$\omega ={}\pm\, \wt_{\mathrm{pl}}-i\gamma$, and when $\gamma \to 0$ 
these poles pass into the poles of the Green function 
$G_{\mathrm{pl}}^{\,r}(\omega)$ (Fig.~\ref{Fig1}, left panel). 
Evidently for these poles there is no need to specify the bypass rules 
when integrating with respect to $\omega$ because,  at positive  
non-zero values of $\gamma$,  these poles are located in the lower 
half-plane of the complex variable $\omega$, certainly outside of the 
real axis. Hence the symbolic equality \eqref{e17d} is not applicable for 
taking into account the contribution of these poles to the imaginary 
part of the Green function $G_{\mathrm{D}}^{\,r}(\omega)$.
In addition this Green function has a `superfluous' pole at the origin  
[the first term in the right hand side of \eqref{e48}] which has no 
prototype in the Green function $G_{\mathrm{pl}}^{\,r}(\omega)$. 
In this case the equality \eqref{e17d} works however its application 
gives contribution to the real part of the Green function 
$G^{\,r}_{\text{D}}(\omega)$ but not to its imaginary part because 
of the imaginary coefficient $(-2i \gamma g/\omega_{\text{pl}}^2)$ 
in the front of the pole factor $(\omega +i \varepsilon)^{-1}$ in the 
term under consideration. It is to be noted that this irregular finite 
contribution emerges for arbitrary, no matter how small, values of the 
constant $\gamma$. Hence this contribution is non-analytic in $\gamma$
 when $\gamma \to 0$. Such an nonanalyticity  has been revealed 
 previously in Ref.\  \cite{BorEPJ}.

This analysis of analytical properties of the Green function 
$G^{\,r}_{\text{D}}(\omega)$ shows that the test of the important 
condition~\eqref{e44a} is impracticable in this case.

From all foregoing it follows uniquely that the Green function with 
the Drude model permittivity, $G_{\mathrm{D}}^{\,r}(\omega)$, cannot 
be used in the exact AC-relation \eqref{e21}.  

However different situation arises in considering  the use of 
$\varepsilon_{\mathrm{D}}(\omega)$ in FDT and consequently in the 
Lifshitz formula.  First of all it is necessary to remind that we 
finally are interested in the possibility of  making use of 
 $\varepsilon_{\mathrm{D}}(\omega)$ 
 for describing  the experimental data on the Casimir forces the 
 accuracy of which is about 1\%. As it was noted in Sec.\ IV the 
 accuracy of FDT in the case of dissipative system may be estimated
by the extent  of dissipative processes, that is  by the ratio  of 
the parameter  responsible for dissipation to the parameter 
characterizing reversible part of dynamics. In the Drude model 
\eqref{e45} such parameters are respectively $2\gamma $ 
and $\omega_ {\text{pl}}$. In our consideration the typical values of 
these parameters are provided by gold~\cite{PRE}
\begin{equation}
\label{e50}
2\gamma =34{.}5~\text{meV},\quad \omega_{\text{pl}}= 9{.}03~\text{eV}
\,{,}
\end{equation}
and the aforesaid ratio is
\begin{equation}
\label{e51}
\frac{2\gamma}{\omega_{\text{pl}}}=3{.}83\times 10^{-4}
\,{.}
\end{equation}
This means that the use  of  $\varepsilon_{\mathrm{D}}(\omega)$ in 
FDT and, consequently in the Lifshitz formula, is quite admissible from 
the practical point of view.

Here it is worthy to recall the following  peculiarity of the Lifshitz formula. 
In its final form this formula contains dielectric permittivity
$\varepsilon (\omega)$ only along the upper imaginary axis $\omega $, 
that is, $\varepsilon (i \zeta)$ with real $\zeta$~\cite[Chap.\ VIII, Eq.\  
(81,9)]{LL9}. As known any physically admissible  phenomenological 
function $\varepsilon(\omega)$ takes on here a real 
values~\cite[Chap.\ IX, \S 82]{LL8}. It is this feature of the Lifshitz 
formula that makes it practically  irreplaceable  in description 
of the experimental data on the Casimir forces.

Plasma model  and Drude model permittivities, Eqs.\ \eqref{e42} and 
\eqref{e45}, assume real values for pure imaginary $\omega$.

Thus we infer that both permittivities 
under consideration can be used  in the Lifshitz formula practically on 
the same physical grounds.  Therefore it is for comparison with 
the experiment  to decide which permittivity is more suitable to  
apply in this formula.

   It is worthy to note that  this inference does not imply that the plasma model and 
 the Drude model permittivities applicable to arbitrary frequencies. The point is these
 models are based on a simple classical presentation of the electron dynamics in
 the  dielectric or metal materials. In the  plasma model electrons are considered
 to be free.  The   Drude model, in addition, takes into account the possibility to  collide
 of such electrons  with the ions of the crystal lattice. The first model is determined 
by one dimensional parameter only,
plasma frequency $\omega_{\text{pl}}$. The latter model involves,
in addition to  $\omega_{\text{pl}}$, the free run of the electron that  leads to finite conductivity 
$\sigma $ or  damping parameter  $\gamma$ (see, for instance, \cite[Chap.\ IV, \S~39]{Sommer}).

 Therefore it is clear from the physical point of view  that the limits of applicability of these 
two models should be.   For example, it is well known 
 that the  plasma model  permittivity cannot be applicable
 at low frequencies because it is derived  in the limit of high frequencies
\cite[Chap.\ IX, \S\   78]{LL8}. 
Theoretically it is reasonable to suggest that the Drude 
model does not work within some frequency range where it was not independently 
tested yet.

Besides these two rather simple models it is also interesting to use  more
complicated models for dielectric permittivity.
In Ref.\ \cite{Galina}  the spatially 
nonlocal dielectric model is considered which takes dissipation into account 
and simultaneously agrees with all the measurement data for the Casimir force. 

  
In addition to the Drude model permittivity  it is appropriate here to consider briefly 
the oscillator dielectric permittivity describing  isotropic dielectric materials
\begin{equation}
\label{Chen}
\varepsilon(\omega)=1+ \sum_{j=1}^{K}\frac{g_j}
{\omega_j^2-\omega ^2-i \gamma_j\omega},
\end{equation}
where $g_j$ are the oscillator strengths, $\omega_j\neq 0$  are 
the oscillator frequencies, and $\gamma_j$ are the dissipation parameters.
In this model the dielectric atoms  are considered to be {\it weakly damped}
harmonic oscillators (see, for example, 
\cite{BW}, \cite[Sec.\ II]{NP}).

The indicated permittivity reduces to the Drude model
only in one particular case of the zero oscillator frequency. For all 
non-zero oscillator frequencies specific for the core electrons in dielectric
materials this permittivity describes the narrow resonances with respective
{\it small dissipation characteristic} for the two-directional exchange of heat 
in the state of thermal equilibrium (as opposed for the Drude model which
possesses the wide resonance at zero oscillator frequency). For this reason,
the oscillator dielectric permittivity leads to an agreement between the
Lifshitz theory and the measurement data for dielectric materials, on the 
one hand, and the Nernst heat theorem, on the other hand (see, e.g., 
F.\ Chen {\it et al.,} Ref.\  \cite{F-Chen}).

Here   the following question  naturally arises,  namely: Is the use of the oscillatory permittivity,
Eq. \eqref{Chen}, in the Lifshitz formula 
theoretically grounded,  like the use of  plasma model and Drude model  permittivities?
The answer to this question should be {\it positive} due to the following reasoning.

This answer   can be obtained in the analogous  way as it has been  done for the Drude model
(see Eqs.\ \eqref{e50}  and \eqref{e51} above),  i.e.,   we have simply to calculate the expression 
\begin{equation}
\label{ratio}
\sum_{j=1}^{K}\frac{g_j \gamma_j}{\omega_j}, \quad j=1, \ldots, K\, {.}
\end{equation}

However this numerical calculation is beyond  the scope of our study 
dealing with the plasma model and Drude model permittivities first of all. 
It is only worthy to note here that,  by definition of the oscillator model 
\eqref{Chen}, the pertinent oscillators are considered to be {\it weakly 
damped}~\cite[Sec.\ II]{NP}.  It implies that the sum \eqref{ratio} should be  small.
 
\section{Conclusion}

In the paper the analysis  of theoretical base 
utilized in derivation of the Lifshitz formula
is conducted thoroughly. It is revealed that in different approaches
to obtain this formula the use of FDT is an essential step.
In view of this  a special approach to deriving  the FDT is
proposed  that shows clear in what way the dissipation may be taken 
into account in this theorem and, consequently, in the Lifshitz formula. 
Proceeding from this it is shown  clear that  there are no substantial 
theoretical arguments in favour of using either plasma mode permittivity
or Drude model permittivity in the Lifshitz formula. 
The decision in this question rests with the comparison 
of theoretical  calculations with the experiment. 

In the course of study  it is explained  how the fluctuation-dissipation
theorem  works in the case of the system with reversible dynamics, 
that is when dissipation is absent. Thereby it is proved that explicit 
assertion according to which this theorem is inapplicable to systems 
without dissipation is erroneous. 

The research is based on making use of the rigorous formalism of equilibrium 
two-time Green functions in statistical physics at finite temperature and real time.
This formalism is grounded on  the Hamiltonian approach and hence can be considered 
as a first principle.

In this connection it is also worth noting  the recent works  where
the dissipation properties in graphene are described on the basis 
of the first principles of QED at non-zero temperature in the framework of the
Dirac model~\cite{BFGV}. The respective results were used to calculate the Casimir force
in the framework of the Lifshitz theory and found to be in excellent agreement 
with the measurement data (please see \cite{LZKMM-Let,LZKMM-PRB})
and with the requirements of thermodynamics \cite{KM-PRD}.

\appendix

\section{Transition to the imaginary frequencies in 
derivation of the Lifshitz formula with plasma model permittivity at finite temperature} 

The imaginary frequencies in the Lifshitz formula at finite 
temperature~$T\neq 0$ can be introduced in the same way as
it has been done in our work~\cite{NP} for~$T=0$. We~explain
here the main points of this transition for the plasma model
at~first.

The Casimir energy at~$T=0$ was defined by us as the spectral 
sum\footnote{The method of spectral summation,
or mode-by-mode summation, is, in fact, the development of
the well known in statistical physics method of natural
(proper) modes, which goes back to the works of Debye~\cite{Debye-1},
Planck~\cite{Planck}, Sommerfeld~\cite{Sommer}. The central point of this  method
lies in the fact that each natural mode of the classical
system is considered to be the quantum oscillator.} 
of the zero point energy of electromagnetic field 
excitations~$\hbar \omega /2$:
\begin{equation}
E(2a)= \frac{\hbar}{2}\sum_{\sigma}\int\frac{d^2 \mathbf{k}}{(2\pi)^2} 
\left [ \sum_{n}\omega_n(\sigma, k, a)+
\int \limits_{\omega_+}^{\infty }\omega 
\Delta \rho (\sigma, \omega, k,a) d\omega \right ] 
- (a \to \infty)\ {.}
\label{b1}
\end{equation}
This energy is per unite area of the boundaries.
In~Eq.~(\ref{b1}) we use the notations which are generally accepted
in the problem in question (see~Sec.~III of our paper~\cite{NP}). 
The~first term in square brackets in Eq.~(\ref{b1}) is the contribution
of the discrete part of the spectrum and the second term represents
the contribution of the continuous part of the electromagnetic 
excitations spectrum in the problem under consideration.
In~Eq.~\eqref{b1} and further \mbox{$\sigma =\text{TE,TM}$}.

The discrete eigenvalues~$\omega _n $ in Eq.~\eqref{b1} are the
frequencies of the surface modes~$\omega_{sm}$ and the waveguide
modes~$\omega_{wg}$. They are given by the real positive roots
of the frequency equations
\begin{equation}
\label{b2}
D_\sigma (\omega, k,a) \equiv 1-r^2_\sigma (\omega) e^{4i a k_2 }=0\, {,}
\end{equation}
where $r_\sigma (\omega)$ is the reflection amplitudes~\cite{NP}
\begin{equation}
\label{b2a}
r_{\text{TE}}=\frac{\mu_2 k_1-\mu_1 k_2}{\mu_2 k_1+ \mu_1 k_2}, 
\quad 
r_{\text{TM}}=\frac{\varepsilon_2 k_1-\varepsilon_1 k_2}{\varepsilon_2 k_1+ \varepsilon_1 k_2}\,{.}
\end{equation}
Just as in Ref.~\cite{NP}, we assume that~$\varepsilon_2=\mu_1=\mu_2=1$ and
the permittivity~$\varepsilon_1(\omega)$ is defined in Eq.~\eqref{e42}. 
The wave vectors $k_1$ and $k_2$ are specified below.

The spectral density shift in Eq.~\eqref{b1}, $\Delta \rho (\omega)$, 
as the function of the real frequency~$\omega$, is given by a known
expression
\begin{equation}
\label{b3}
\Delta \rho(\omega) =\frac{1}{\pi}\, 
\frac{d}{d \omega }\delta(\omega)\, {,}
\end{equation}
where~$\delta (\omega)$ is the phase shift. Thus Eq.~\eqref{b1}
involves only real positive frequencies~$\omega$.

\begin{figure*}[t]
\centerline{\includegraphics[width=0.98\textwidth]{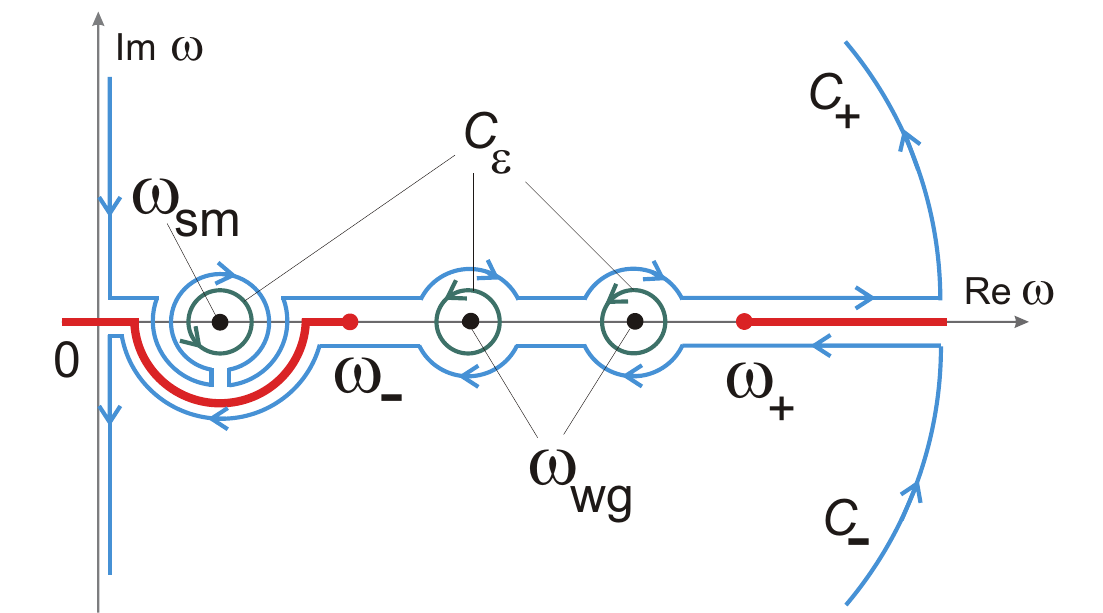}}
\caption{(Color online) The contours $C_{+}$ and $C_{-}$ in 
the complex $\omega$ plane which are used when going on to 
the imaginary frequencies. For simplicity, the spectrum 
possessing one surface mode $\omega_{\text{\,sm}}$, 
two waveguide modes $\omega_{\text{\,wg}}$, and 
the continuous part $\omega > \omega_{+}$ is considered. 
The cuts starting at the points $\omega_{-}$ and $\omega_{+}$
are shown by thick lines.}
\label{fig2}
\end{figure*}

In order to pass to imaginary frequencies the outcome to complex
plane~$\omega$ is needed. For~this aim in Ref.~\cite{NP} two cuts 
have been made on this plane, namely, the first cut between the
points~$-\omega_{-}(k)$ and~$\omega_{-}(k)$ and the second one 
on the interval~$(\omega_{+}(k),\infty)$ [see~Fig.~\ref{fig2}]. 
The points~$\omega_{-}(k)$ and~$\omega_{+}(k)$ are the branch 
points of the functions
\begin{gather}
ck_1(\omega)=\sqrt{\omega^2-\omega_+^2(k)},
\quad
ck_2(\omega)=\sqrt{\omega^2-\omega_-^2(k)}, \label{b3a}
\\
\omega_+^2(k)=\omega_{\mathrm{pl}}^2+c^2k^2, \quad \omega_-^2(k)=c^2k^2, 
\label{b3b}
\end{gather}
and, at the same time, they determine the boundaries between the
different branches of the spectrum, namely, on the interval~$0 
< \omega_{sm}(k) < \omega_{-}(k)$ the surface mode frequencies
are located, on the interval~$\omega_{-}(k) < \omega_{wg}(k) 
< \omega_{+}(k)$ the waveguide mode frequencies lie, 
and the interval~$\omega_{+}(k) < \omega(k) < \infty$ belongs
to the continuous spectrum.

The single-valued branches of the functions~\eqref{b3a} are extracted 
by the requirements that the function~$k_2(\omega)$ takes on real
positive values on the upper edge of the cut~$(\omega_+(k), \infty)$, 
and the values of the function~$k_1(\omega)$ on the upper edge of the
cut~$(-\omega_{-}(k), \omega_{-}(k))$ have the argument~$\pi/2$. 
Taking account of this one can easily be convinced that the 
functions~$D_\sigma(\omega)$ and~$\overline{D_\sigma(\omega)}$ 
acquire the same values in
 the interval~$(\omega_-(k),\omega_+(k))$ 
of the real axis and at the opposite points lying on the different
edges of the cut~$(0,\omega_{-}(k))$.

Further we shall need the properties of the frequency equation~\eqref{b2}, 
that is of the function~$D_\sigma(\omega)$, in the upper half-plane~$\Omega$ 
of the complex variable~$\omega$. The plasma model permittivity~\eqref{e42} 
is considered.

\begin{enumerate}

\item \label{i1}
The function~$D_\sigma(\omega)$ tends to~$1$ 
for~$|\omega| \to \infty,\; \omega \in \Omega$. 

\item\label{i2}
$D_\sigma(\omega)=D_\sigma(-\omega)$ see Eqs.~\eqref{b2}, \eqref{b2a},
\eqref{b3a}, \eqref{b3b}, and~\eqref{e42}.

\item\label{i3}
Along  the imaginary positive semi-axis the function~$D_{\sigma}(\omega)$ 
assumes real values.
    
\item\label{i4} 
Equation~\eqref{b2} has no complex roots in~$\Omega$, that is 
the frequencies of the quasi-normal modes, if they exist, have 
\textit{negative imaginary part}. Otherwise the macroscopic 
electrodynamics would admit the natural modes increasing in time
without limits. Recall that the time dependence is taken,
as usual, in the form~$e^{-i\omega t}$. These properties of the
quasi-normal modes are explicitly ascertained for electromagnetic
oscillations of a perfectly conducting sphere~\cite{JPA06} and a
dielectric ball without dispersion~\cite{Debye, Gast, shape}.
       
\item\label{i5} 
As follows from Eq.~\eqref{b2}, the poles of the 
function~$D_{\sigma}(\omega)$ may be generated only
by the reflection amplitude, $r_{\sigma}(\omega)$, 
and their position on the complex $\omega$-plane is
independent of the gap width~$2 a$. Therefore the
contribution of these poles to the vacuum energy~\eqref{b1} 
will be canceled when subtracting in this formula the
respective limiting expression obtained for~$a\to \infty$. 
Hence these poles can be disregarded.
               
\item \label{i6} 
The function~$\overline{D_\sigma(\omega)}$ possesses the same
properties in~$\overline{\Omega}$ .

\end{enumerate}

Employing the argument principle theorem (see Sec.~3.4 of Ref.~\cite{tit}) 
we represent the contribution of the discrete spectrum into~\eqref{b1} 
in the following form
\begin{equation}
\label{b3c}
\omega_n=\frac{1}{2\pi i}  \oint_{C_\varepsilon} d\omega \,\omega 
\frac{d}{d\omega} \ln  D_\sigma(\omega)\,{,}
\end{equation}
where~$C_\varepsilon$ is a circle of the radius~$\varepsilon$ 
with~$\varepsilon \to 0$, see Fig.~2. The spectral density
shift~$\Delta \rho$, as the function of the complex variable~$\omega$, 
is defined ultimately by the function~$D_{\sigma}(\omega)$
\begin{equation}
\label{b4}
\Delta \rho (\omega)=\frac{1}{2\pi i}\, \frac{d}{d\omega}
\ln \frac{\overline{D_\sigma(\omega+i0)}}{D_\sigma(\omega+i0)}
\end{equation}
(see details in Ref.~\cite{NP}). In Eq.~\eqref{b4} we explicitly note
that the function~$D(\omega+i0)$ is calculated at the upper edge of the
cut~$(\omega_{+},\infty)$ and the function~$\overline{D(\omega+i0)}$ 
is calculated at the lower edge of this cut.

Taking into account Eqs.~\eqref{b3c} and~\eqref{b4}, we can now represent 
the vacuum energy~\eqref{b1} in the following form
\begin{equation}
\label{b4a}
E(2a)=\frac{\hbar}{2}\sum_\sigma\int_0^\infty 
\frac{k dk}{2\pi} \frac{1}{2\pi i}
\left[
\sum_{C_\varepsilon}\oint_{C_\varepsilon} d\omega \,\omega 
\frac{d}{d\omega}\ln D_\sigma(\omega) +
\int_{\omega_+}^\infty d\omega\,\omega \frac{d}{d\omega} 
\ln \frac{\overline{D_\sigma(\omega+i0)}}{D_\sigma(\omega+i0)}
\right]\!.
\end{equation}

The properties of the functions~$D_{\sigma}(\omega)$ 
and~$\overline{D_{\sigma}(\omega)}$ enumerated above 
enable us to write the following equalities
\begin{gather}
\frac{1}{2\pi i} \oint_{C_+}\omega \frac{d}{d \omega } 
\ln D_\sigma(\omega) d\omega =0\,{,} \label{b5}\\
\frac{1}{2\pi i} \oint_{C_-}\omega \frac{d}{d \omega } 
\ln \overline{D_\sigma(\omega)} d\omega =0\,{,} \label{b6}
\end{gather}
where the contours~$C_+$ and~$C_-$ enclose, respectively, 
the first and the fourth quadrants of the $\omega$-plane
as depicted in Fig.~2.

Further we proceed in the following way: the integral in Eq.~\eqref{b4a},
containing the function~$D_{\sigma}(\omega+i0)$ on the upper edge of the
cut~$(\omega_{+}, \infty)$, we express from Eq.~\eqref{b5} and the integral
including the function~$\overline{D_{\sigma}(\omega+i0)}$, evaluated on the
lower edge of this cut, we express from Eq.~\eqref{b6}. Doing in this way
we can disregard, in virtue of the property~i), the contributions due to
the arcs of the big circle. One can easily see from Fig.~2 that along the 
contours $C_\varepsilon$ the contributions of the discrete spectrum and 
the continuous spectrum are mutually canceled. In addition, the 
integrals with the function~$D_{\sigma}(\omega+i0)$ 
and~$\overline{D_{\sigma}(\omega+i0)}$ between the origin and the 
point~$\omega_{+}$ are reciprocally canceled~too. As the result the 
integration only along the imaginary axis~($\omega=i\zeta$) survives:
\begin{equation}
\label{b7}
E(2a)=-\frac{\hbar}{2}\sum\limits_\sigma\int_0^\infty\frac{k dk}{(2\pi)^2}
\left[
\int_{0}^{\infty} d\zeta\, \zeta \frac{d}{d\zeta} \ln D_\sigma(i\zeta,k)+
\int_{-\infty}^0 d\zeta\, \zeta \frac{d}{d\zeta}
\ln \overline{D_\sigma(i\zeta,k)}
\right]\!{.}
\end{equation}
Taking into account the property~\ref{i3}), we can join two integrals
in Eq.~\eqref{b7} in one integral with the limits~$(-\infty, \infty)$
and after that we accomplish the integration by parts. The terms outside 
the integral vanish in view of the property~\ref{i1}). As~a result
Eq.~\eqref{b7} acquires the form
\begin{equation}
\label{b8}
E(2a)=\frac{\hbar}{2}\sum_\sigma \int_0^\infty \frac{k\,dk}{(2\pi)^2}
\int_{-\infty}^\infty d\zeta\,\ln D_\sigma (i \zeta,k)=
\frac{\hbar}{2\pi}\sum_\sigma \int_0^\infty \frac{k\,dk}{2\pi}
\int_{0}^\infty d\zeta\,\ln D_\sigma (i \zeta,k)
\end{equation}
(see Eqs.~(71) and~(72) in our paper~\cite{NP}). The last equality in
Eq.~\eqref{b8} is obtained on account of the property~\ref{i2}).

Differentiation of the vacuum energy~\eqref{b8} with respect to the
gap width~$2a$ results in the standard representation of the Lifshitz
formula at zero temperature (see Eqs.~(73) and~(75) in Ref.~\cite{NP}).

We shall not write out these well known formulas but at once go over
to derivation of the Lifshitz formula at finite temperature~$T$. 
To~this end the free energy~$\mathcal{F}$ of electromagnetic field
in the problem at hand should be found. We~accomplish this task
by summing up, with respect to the spectrum of electromagnetic 
excitations, the free energy of quantum oscillator
\begin{equation}
\label{b9}
f(\omega)=\frac{\hbar \omega}{2}+ k_{\text{B}}T \ln \left[
1- \exp\left(-\frac{\hbar \omega}{k_{\text{B}}T}\right)   
\right]\!{,}
\end{equation}
where $k_{\text{B}}$ is the Boltzman constant. Thus in all formulas
beginning with Eq.~\eqref{b1}, the zero point energy~$\hbar\omega/2$
should be replaced by~$f(\omega)$. The~function $f(\omega)$ has
logarithmic branch points for~$\omega=i \zeta_m$, where~$\zeta_m$ are
the Matsubara frequencies
\begin{equation}
\label{b9a}
\zeta_m=2\pi m \kb T/\hbar,
\quad
m=0,\pm 1, \pm 2, \ldots \,{.}
\end{equation}
The introduction of respective cuts in the $\omega$-plane can be avoided
by using formal and nevertheless rather rigorous method~\cite{NP-JMP, NPerW},
namely, the function~$f(\omega)$ in Eq.~\eqref{b9} should be represented
by the series\footnote{As it will be shown further the branch points of
the function~$f(\omega)$ in Eq.~\eqref{b9}, $\omega=i \zeta_m$, result
in singular contributions proportional to~$\delta(\zeta-\zeta_m$).}
\begin{equation}
\label{b10}
f(\omega)=\frac{\hbar \omega}{2} 
-k_{\text{B}}T\sum_{m=1}^{\infty}\frac{1}{m}\exp\left(
-\frac{m\hbar\omega}{k_{\text{B}}T}
\right)\!{.}
\end{equation}
It is easy to see that after substitution of~$\hbar\omega/2$ 
for~$f(\omega)$ in Eq.~\eqref{b7} the expression
\begin{equation}
\label{b11}
\frac{f(i\zeta)}{i}\equiv\varphi (\zeta)=
\frac{\hbar \zeta}{2}+i k_{\text{B}}  
T\sum_{m=1}^{\infty}\frac{1}{m}\exp\left(
-i\frac{m\hbar\zeta}{k_{\text{B}}T}
\right){.}
\end{equation}
will be here in place of~$\zeta$. In this notation Eq.~\eqref{b7} 
assumes the form
\begin{equation}
\label{b12}
\mathcal{F}(2a)=-\sum_\sigma\int_0^\infty 
\frac{k dk}{(2\pi)^2} \left[
\int_0^\infty d\zeta \varphi(\zeta)\frac{d}{d\zeta}
\ln D_\sigma (i\zeta,k)+
\int_{-\infty}^0 d\zeta \varphi(\zeta)\frac{d}{d\zeta}
\ln \overline{ D_\sigma (i\zeta,k)}
\right]\!{.}
\end{equation}
Taking into account the property~\ref{i3}), we can again join two
integrals in Eq.~\eqref{b12} into~one
\begin{equation}
\label{b13}
\mathcal{F}(2a)=-\sum_\sigma\int _0^\infty 
\frac{k dk}{(2\pi)^2}\int_{-\infty}^\infty d\zeta \,\varphi(\zeta)
\frac{d}{d\zeta}\ln D_\sigma(i\zeta,k)\,{.}
\end{equation}
The integration by parts with allowance for the property~\ref{i1}) gives
\begin{align}
\lefteqn{  \mathcal{F}(2a)=\sum_\sigma \int_0^\infty\frac{kdk}{(2\pi)^2} \int_{-\infty}^{\infty}d\zeta\,\varphi'(\zeta) \ln D_\sigma (i\zeta,k) }\nonumber \\
&= \hbar \sum_\sigma \int_0^\infty\frac{kdk}{(2\pi)^2}\int_{-\infty}^\infty d\zeta \left [
  \frac{1}{2}+\sum_{m=1}^\infty \exp\left (
  -i\mz
  \right )
  \right ] \ln D_\sigma (i\zeta,k) \nonumber \\
  &=  \hbar \sum_\sigma \int_0^\infty\frac{kdk}{(2\pi)^2}\int_{-\infty}^\infty d\zeta \left [
\left ( \frac{1}{2}+\sum_{m=1}^\infty \cos
\mz
  \right )-i \sum_{m=1}^\infty \sin \mz
  \right ] \ln D_\sigma (i\zeta,k)\,{.} \label{b14}
\end{align}
By~virtue of the property~\ref{i2}) the second sum over~$m$ in 
Eq.~\eqref{b14} does not contribute to the integral over~$d\zeta$.
Now~we take advantage of the Fourier series representation for the
``comb'' of $\delta$-functions\footnote{Equation~\eqref{b15} expresses
the following fact. The~function $\pi \kb T\delta(\zeta)$, given at first
on the expansion interval $(-\pi \kb T/\hbar,\pi \kb T/\hbar)$, 
generates the Fourier series in the right hand side of Eq.~\eqref{b15}. 
This series, in its turn, extends the function $\pi \kb T\delta(\zeta)$ 
to the whole infinite line $-\infty < \zeta < \infty$ with the period
$2 \pi \kb T/\hbar$. Just this is stated in the left hand side of
Eq.~\eqref{b15} (see, for example, 
Ref.~\cite[Chap.~4,  Sec.~4.11]{KK}).}
\begin{equation}
\label{b15}
\pi \kb T \sum _{m=-\infty}^\infty \delta(\zeta-2\pi m \kb T/\hbar)=
\hbar \left(
\frac{1}{2} +\sum_{m=1}^\infty \cos \mz
\right)\!{.}
\end{equation}
The substitution of~\eqref{b15} into~\eqref{b14} replaces the integration
over~$d\zeta$ by the summation over the Matsubara frequencies~\eqref{b9a}
\begin{align}
\mathcal{F}(2a)&=\pi \kb T  \sum_\sigma \int_0^\infty 
\frac{k \, dk}{(2\pi)^2}\sum_{m=-\infty}^{\infty} 
\ln D_\sigma(i\zeta_m,k)\nonumber \\
&= \kb T  \sum_\sigma \int_0^\infty \frac{k \, dk}{2\pi}   
\mathop{{\sum}'}_{m=0}^\infty \ln D_\sigma(i\zeta_m,k)\,{.}
\label{b16}
\end{align}
The primed sign of the sum implies that the term with $m=0$ should be 
multiplied by~$1/2$. The last expression in~\eqref{b16} is obtained 
owing to the property~\ref{i2}). The free energy~\eqref{b16} coincides 
exactly with Eq.~(12.66) in the book~\cite{book}.

Differentiation of Eq.~\eqref{b16} with respect to the width of 
the gap~$2a$, gives at once the Lifshitz formula for the Casimir 
force at nonzero temperature (see Eq.~(12.70) in Ref.~\cite{book}).

The change of the integration variable~$k$ in Eq.~\eqref{b16}
\begin{equation}
\label{b17}
k^2=\frac{\zeta_m^2}{c^2}(p^2-1), 
\quad 
k\,dk=\frac{\zeta_m}{c^2}p\,dp, 
\quad 
1<p<\infty
\end{equation}
enables one to reproduce the Lifshitz formula in the original form 
(see Eq.~(81.9) in Ref.~\cite{LL9} for two identical material 
semi-spaces separated by a vacuum gap).

The transition to the Matsubara frequencies in deriving the Lifshitz
formula, presented above, substantially relied on using the
properties~\ref{i1})--\ref{i6}) of the functions $D_\sigma(\omega, k,a)$ 
in the frequency equations \eqref{b2}.

\begin{acknowledgments}
The author thanks A.L.~Kuzemsky for elucidating the FDT,
I.~Brevik for providing the copy of the Van Kampen paper,
M.~Bordag for very valuable discussions of the problem
touched in the article, and I.G.~Pirozhenko for preparing~Fig.~2.

The author is grateful to anonymous referee for the reports
which promoted more clear presentation and clarification of the obtained results.

 At the initial stage, this work was partially supported by the 
 Heisenberg-Landau Program.
\end{acknowledgments}

\end{document}